\documentclass{PoS}
\usepackage{multirow}
\usepackage[numbers,sort&compress]{natbib}
\newcommand{\VZEROA}       {\rm{VZERO-A}}

\newcommand{\pip}          {$\pi^{+}$}
\newcommand{\pim}          {$\pi^{-}$}
\newcommand{\kap}          {K$^{+}$}
\newcommand{\kam}          {K$^{-}$}
\newcommand{\pbar}         {$\rm\overline{p}$}
\newcommand{\kzero}        {\ensuremath{{\rm K}^{0}_{S}}}
\newcommand{\vzero}        {\ensuremath{{\rm V}^0}}
\newcommand{\lmb}          {\ensuremath{\Lambda}}
\newcommand{\almb}         {\ensuremath{\bar{\Lambda}}}
\newcommand{\allpart}      {$\pi^{\pm}$, K$^{\pm}$, \kzero, p(\pbar) and \lmb(\almb)}

\newcommand{\pp}           {pp}

\newcommand{\PbPb}         {\mbox{Pb--Pb}}
\newcommand{\pPb}          {\mbox{p--Pb}}

\newcommand{\dNdeta}       {\ensuremath{\mathrm{d}N_\mathrm{ch}/\mathrm{d}\eta}}

\newcommand{\s}            {\ensuremath{\sqrt{s}}}
\newcommand{\pt}           {\ensuremath{p_{\rm T}}}
\newcommand{\hlab}         {\ensuremath{\eta_{\rm lab}}}
\newcommand{\ynn}         {\ensuremath{y_{\rm NN}}}
\newcommand{\ycms}         {\ensuremath{y_{\rm CMS}}}

\newcommand{\ppi}          {\ensuremath{{\rm p}/\pi}}
\newcommand{\kpi}          {\ensuremath{{\rm K}/\pi}}

\newcommand{\snn}          {\ensuremath{\sqrt{s_{\rm NN}}}}

\newcommand{\gevc}         {\ensuremath{{\rm GeV}/c}}

\title{Transverse momentum distribution of charged particles and identified hadrons in p--Pb collisions at the LHC with ALICE}

\ShortTitle{Transverse momentum distribution of identified hadrons in p--Pb collisions}

\author{\speaker{Roberto Preghenella} for the ALICE Collaboration\\
  Centro Studi e Ricerche e Museo Storico della Fisica ``Enrico Fermi'', Rome, Italy\\
  Sezione INFN, Bologna, Italy\\
  E-mail: \email{roberto.preghenella@bo.infn.it}}

\abstract{Hadron production has been measured at mid-rapidity by the ALICE experiment at the LHC in proton-lead (p--Pb) collisions at $\sqrt{s_{\rm NN}}$ = 5.02 TeV. The transverse momentum ($p_{\rm T}$) distribution of primary charged particles and of identified light-flavoured hadrons ($\pi^{\pm}$, K$^{\pm}$, K$^{0}_{\rm S}$, p, $\bar{\rm p}$, $\Lambda$, $\bar{\Lambda}$) are presented in this report. Charged-particle tracks are reconstructed in the central barrel over a wide momentum range. Furthermore they can be identified by exploiting specific energy loss (d$E$/d$x$), time-of-flight and topological particle-identification techniques. Particle-production yields, spectral shapes and particle ratios are measured in several multiplicity classes and are compared with results obtained in Pb--Pb collisions at the LHC.

The measurement of charged-particle transverse momentum spectra and nuclear modification factor R$_{\rm pPb}$ indicates that the strong suppression of high-$p_{\rm T}$ hadrons observed in Pb--Pb collisions is not due to initial-state effects, but it is rather a fingerprint of jet quenching in hot QCD matter. The systematic study of the hadronic spectral shapes as a function of the particle mass and of particle ratios as a function of charged-particle density provides insights into collective phenomena, as observed in Pb--Pb collisions. Similar features, that could be present in high-multiplicity p--Pb collisions, will also be discussed.}

\FullConference{The European Physical Society Conference on High Energy Physics -- EPS-HEP2013\\
  18-24 July 2013\\
  Stockholm, Sweden}

\begin{document}

\section{Introduction}

High-energy heavy-ion (AA) collisions offer a unique possibility to study
hadronic matter under extreme conditions, in particular the deconfined
quark-gluon plasma which has been predicted by quantum chromodynamics
(QCD)~\cite{Cabibbo:1975ig, Shuryak:1978ij, McLerran:1980pk,
  Laermann:2003cv}. The interpretation of the results depends
crucially on the comparison with results from smaller collision
systems such as proton-proton (\pp) or proton-nucleus (pA).
Proton-nucleus (pA) collisions are intermediate between
proton-proton (pp) and nucleus-nucleus (AA) collisions both in terms of
system size and number of produced particles. Comparing particle
production in pp, pA, and AA reactions is frequently used to
separate initial state effects, connected to the use of nuclear
beams or targets, from final state effects, connected to the presence of hot and
dense matter. Moreover, pA collisions allow for the
investigation of fundamental properties of QCD; the \pt\ distributions and yields of particles of
different mass at low and intermediate momenta of \pt\ $\lesssim$ 3
\gevc\ (where the vast majority of particles is produced) can provide
important information about the system created in high-energy hadron
reactions.

Previous results on identified particle production in pp and \PbPb\
collisions at the LHC have been reported in~\cite{Aamodt:2011zj,
  Aamodt:2011zz, Abelev2012309, prl-spectra,
  Abelev:2013vea,Chatrchyan:2012qb,Khachatryan:2011tm}. Results on transverse momentum distribution and nuclear modification factor of charged particles in \pPb\ collisions at \snn~=~5.02~TeV have been reported in~\cite{ALICE:2012mj}. In this paper
we report on the measurement of \allpart\ production in \pPb\
collisions at a nucleon-nucleon center-of-mass energy
\snn~=~5.02~TeV.

\section{Sample and Data analysis}

The results presented here were obtained from a sample of the data
collected during the LHC \pPb\ run at \snn~=~5.02~TeV 
in the beginning of 2013. 
Due to the asymmetric beam energies for the proton and lead beams,
 the nucleon-nucleon center-of-mass system 
was moving in the laboratory frame with a rapidity of \ynn\ = $-0.465$ 
in the direction of the proton beam. A detailed description of the ALICE apparatus can be found in~\cite{Aamodt:2008zz} and a description of the data-taking and trigger setup in minimum-bias trigger in~\cite{ALICE:2012xs}.
In order to study the multiplicity dependence, the selected event
sample was divided into seven event classes, based on cuts applied on the
total charge deposited in the \VZEROA\ scintillator hodoscope ($2.8 < \hlab < 5.1$, Pb beam direction).

The ALICE central-barrel tracking covers the full azimuth
within $| \hlab |<0.9$.  The tracking detectors are located inside a solenoidal magnet
providing a magnetic field of 0.5 T. The innermost barrel detector is
the Inner Tracking System (ITS).  The Time Projection Chamber (TPC), the main
central-barrel tracking device, follows outwards.  Finally the
Transition Radiation Detector (TRD) extends the tracking farther away
from the beam axis. 
Charged-hadron identification in the central barrel was performed with
the ITS, TPC~\cite{Alme:2010ke} and Time-Of-Flight
(TOF)~\cite{Akindinov:2013tea} detectors~\cite{performance-paper}.
Three approaches were used for the identification of $\pi^{\pm}$, K$^{\pm}$, and p($\bar{\rm p}$), called ``ITS standalone'', ``TPC/TOF'' and ``TOF
fits'' and are described in details in~\cite{prl-spectra, Abelev:2013vea}. 
Contamination from
secondary particles was subtracted with a data-driven approach, based
on the fit to the transverse distance-of-closest approach to the
primary vertex (DCA$_{xy}$)
distribution with the expected shapes for primary and secondary
particles~\cite{prl-spectra, Abelev:2013vea}.

The \kzero\ and \lmb(\almb) particles were identified exploiting their
``\vzero'' weak decay topology in the channels $\kzero \to \pi^{+}
\pi^{-}$ and $\lmb(\almb) \to \rm{p} \pi^{-} (\rm{\bar{p}} \pi^{+})$.
The selection criteria used to
define two tracks as \vzero\ decay candidates are detailed in~\cite{Aamodt:2011zz, ALICE:2013xaa}.  The contribution from weak decays
of the charged and neutral $\Xi$ to the \lmb(\almb) yield has been
corrected following a data-driven approach.

The study of systematic uncertainties follows the analysis described
in~\cite{prl-spectra, Abelev:2013vea, Aamodt:2011zz, ALICE:2013xaa}
and was repeated for the different multiplicity
bins in order to separate the sources of uncertainty which are 
dependent on multiplicity and uncorrelated across different
bins (depicted as shaded boxes in the figures).

\section{Results}

\begin{figure}[p]
  \centering
  \includegraphics[width=0.495\textwidth]{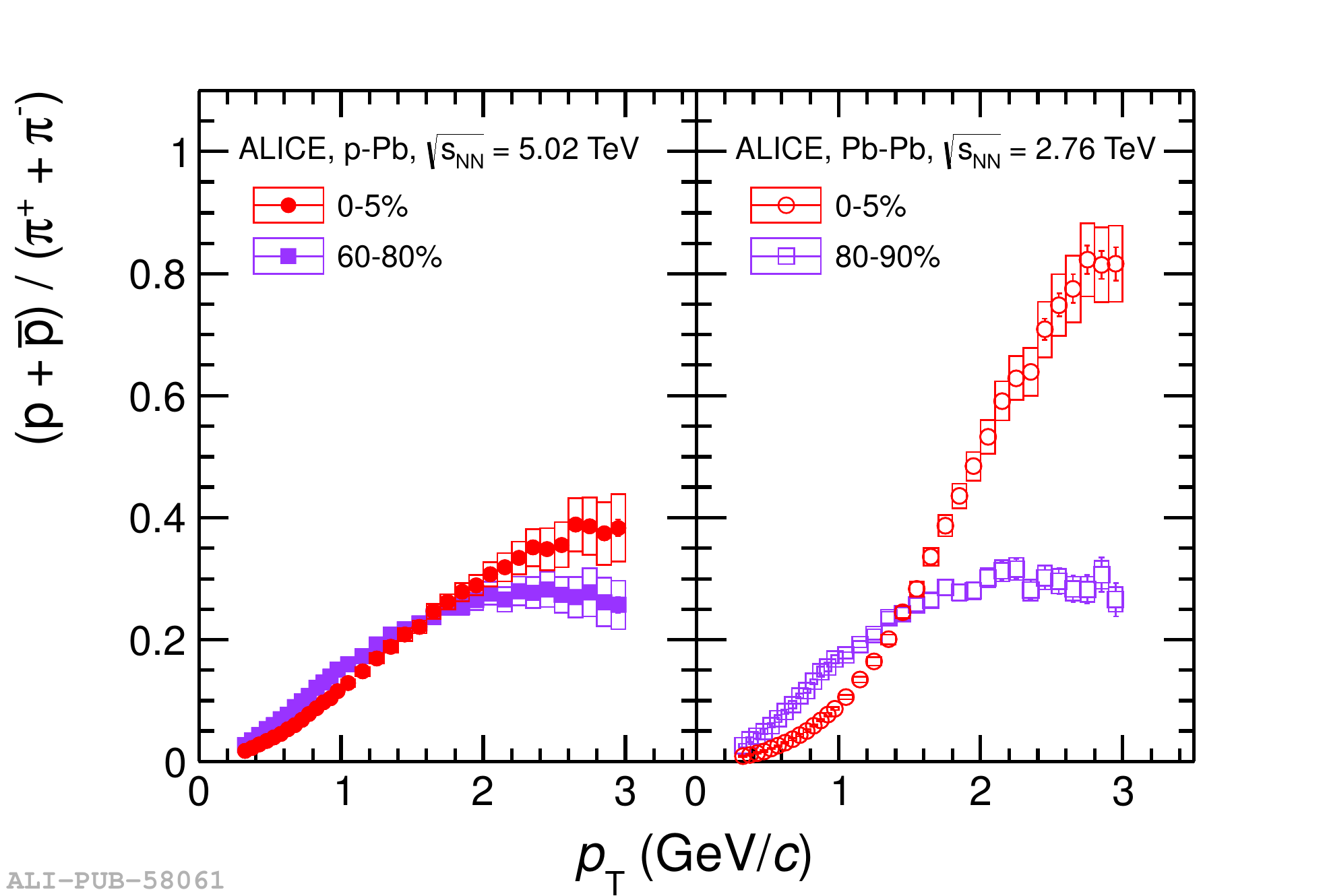}
  \includegraphics[width=0.495\textwidth]{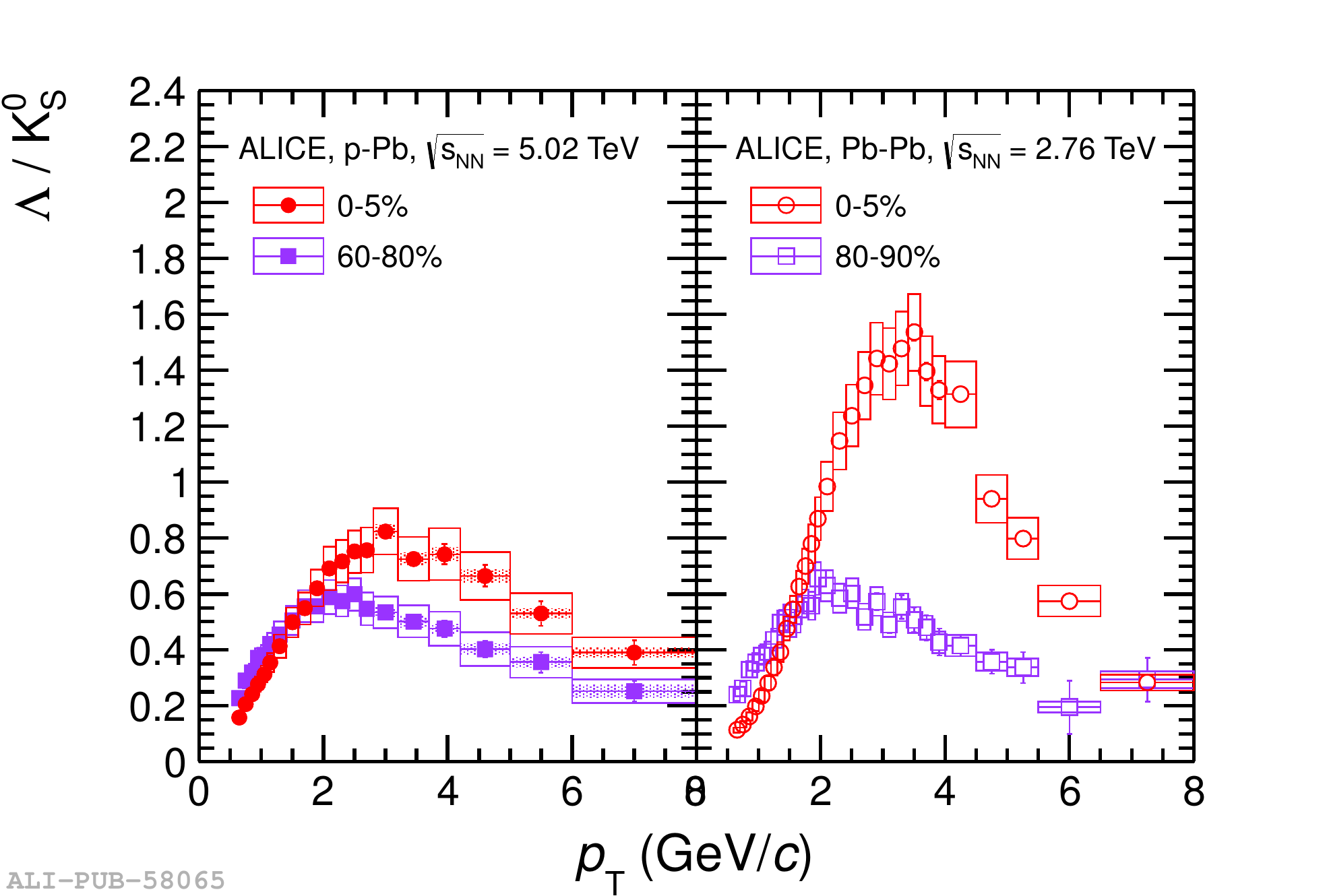}
  \caption{Ratios p/$\pi$ (left) and $\Lambda$/K$_{\rm S}^{0}$ (right) as a function of $p_{\rm T}$ in two multiplicity bins compared to results in Pb--Pb collisions. The empty boxes show the total systematic uncertainty; the shaded boxes indicate the contribution uncorrelated across multiplicity bins (not estimated in Pb--Pb).}
  \label{fig:ratios}
\end{figure}
\begin{figure}[p]
  \centering
  \includegraphics[width=0.495\textwidth]{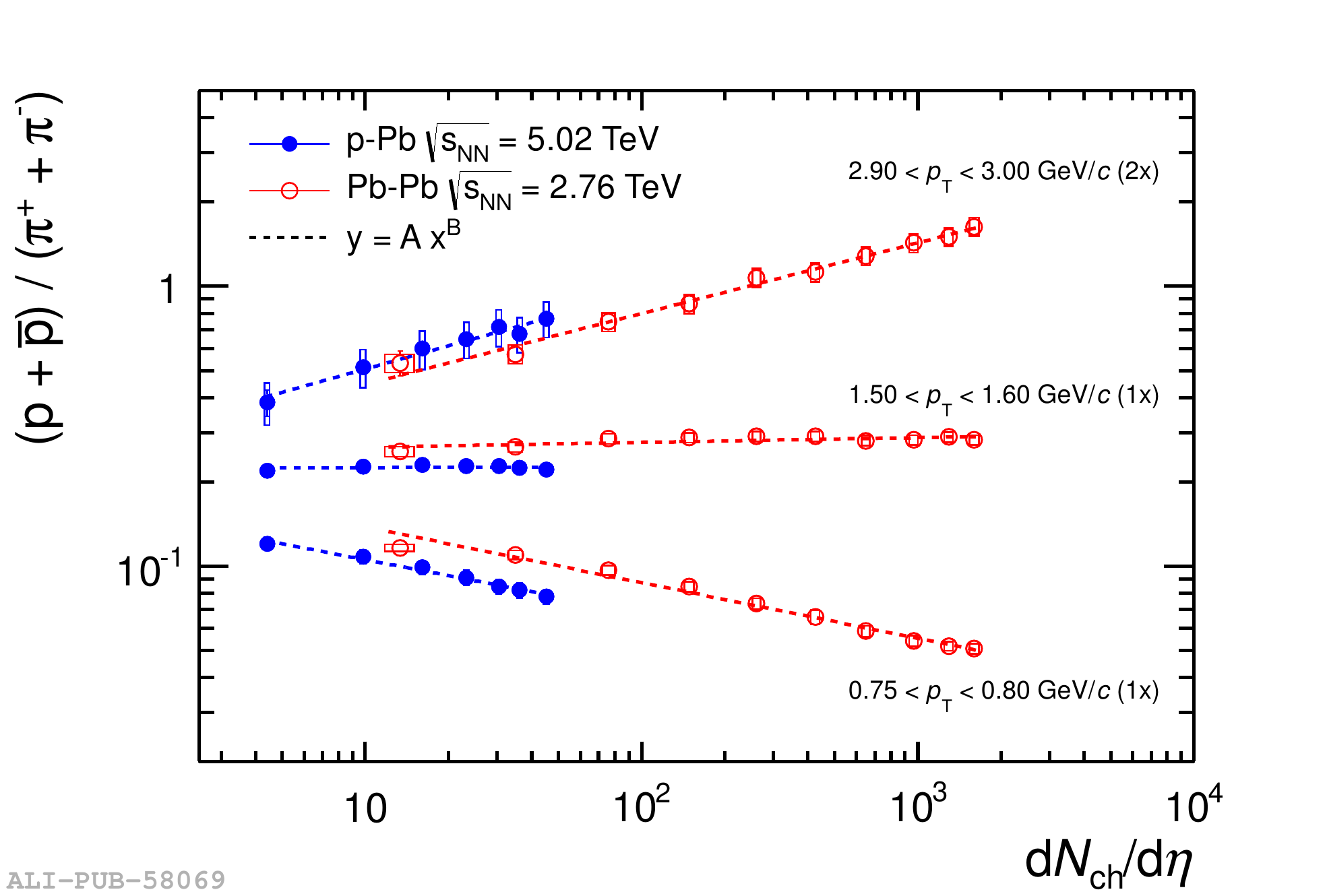}
  \includegraphics[width=0.495\textwidth]{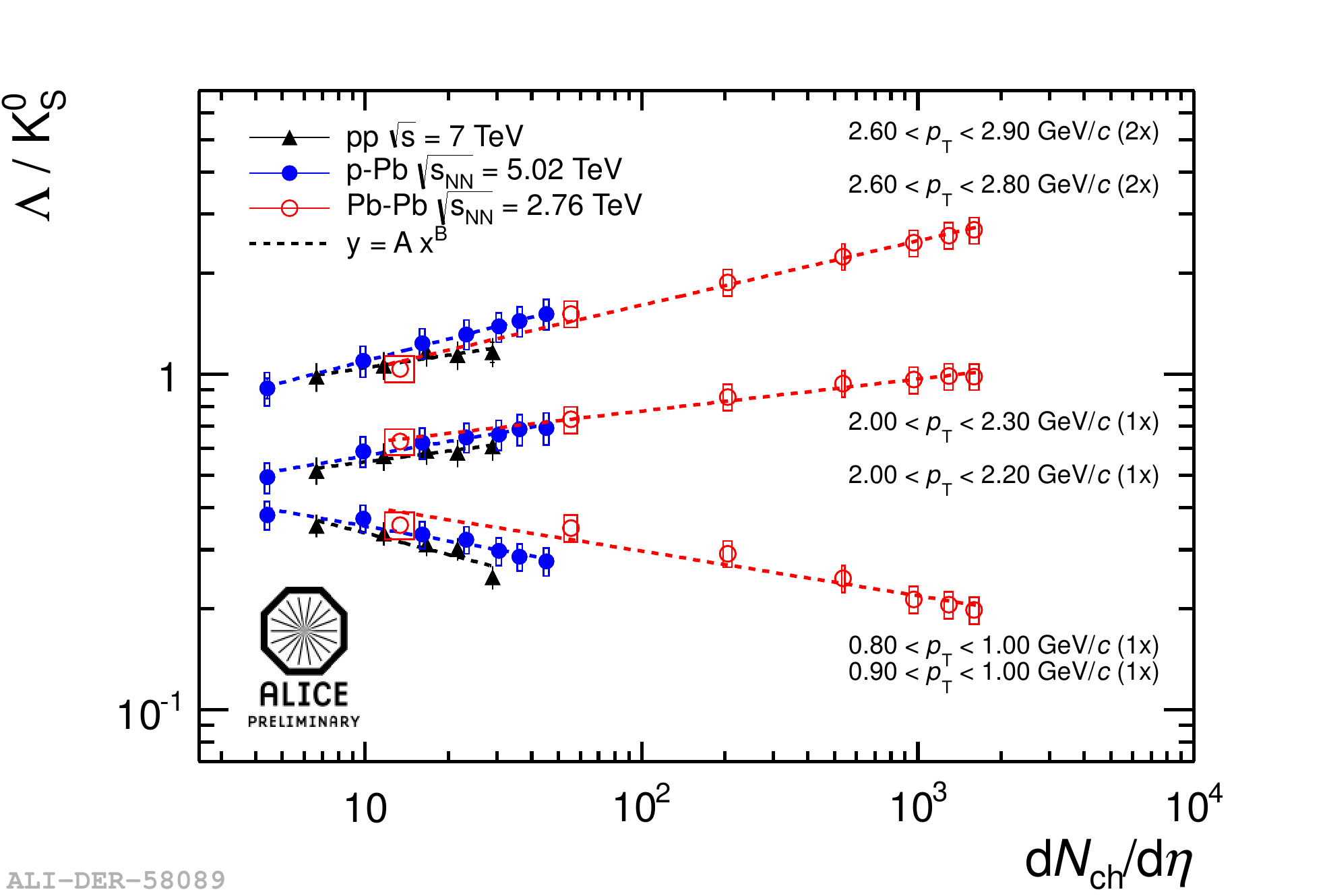}
  \caption{Ratios p/$\pi$ (left) and $\Lambda$/K$_{\rm S}^{0}$ (right) as a function of the charged-particle density d$N_{\rm ch}$/d$\eta$ in three $p_{\rm T}$ intervals in p--Pb, Pb--Pb and pp collisions (pp only shown for $\Lambda$/K$_{\rm S}^{0}$). The dashed lines show the corresponding power-law fit.}
  \label{fig:scaling}
\end{figure}
\begin{figure}[p]
  \centering
  \includegraphics[width=0.495\textwidth]{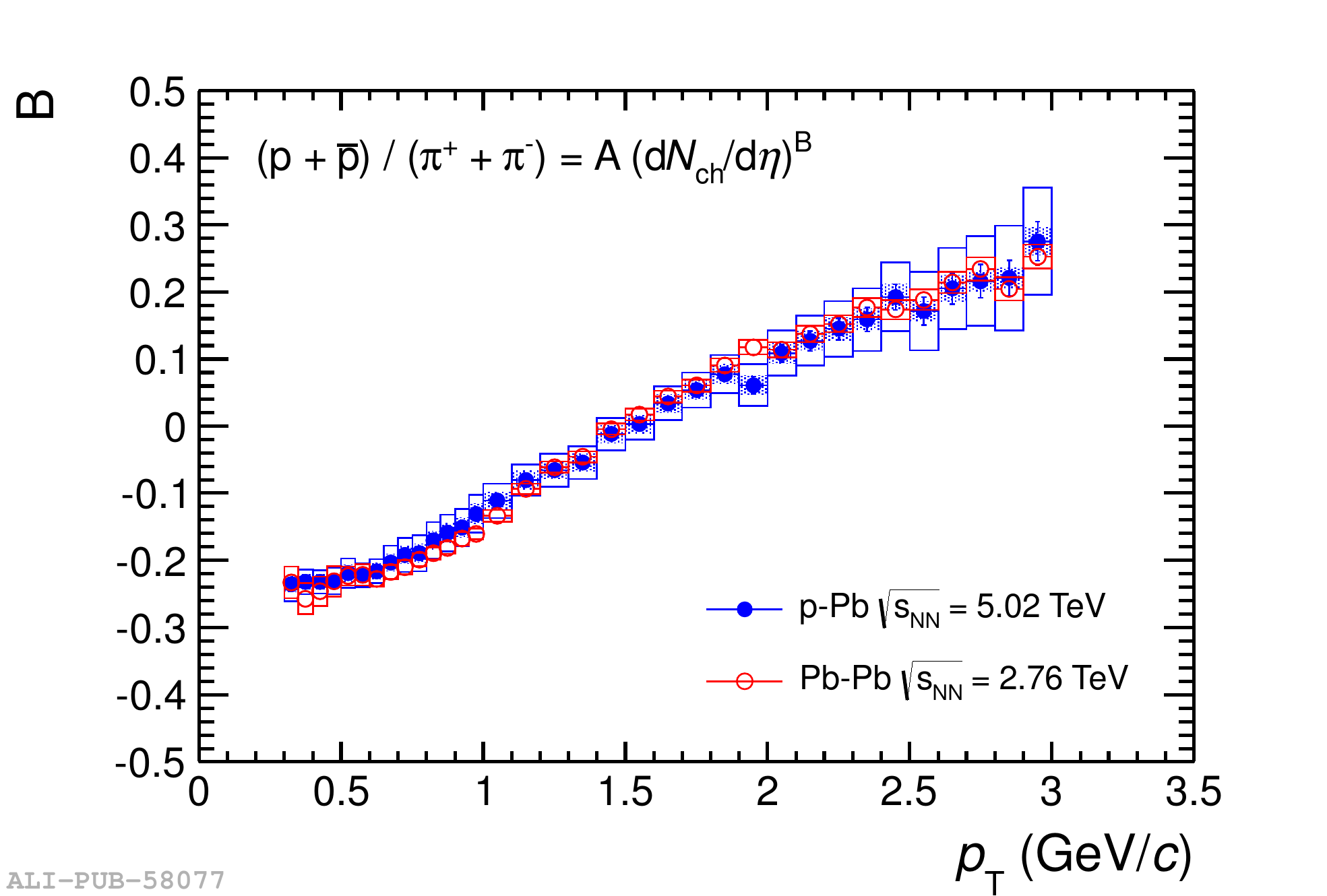}
  \includegraphics[width=0.495\textwidth]{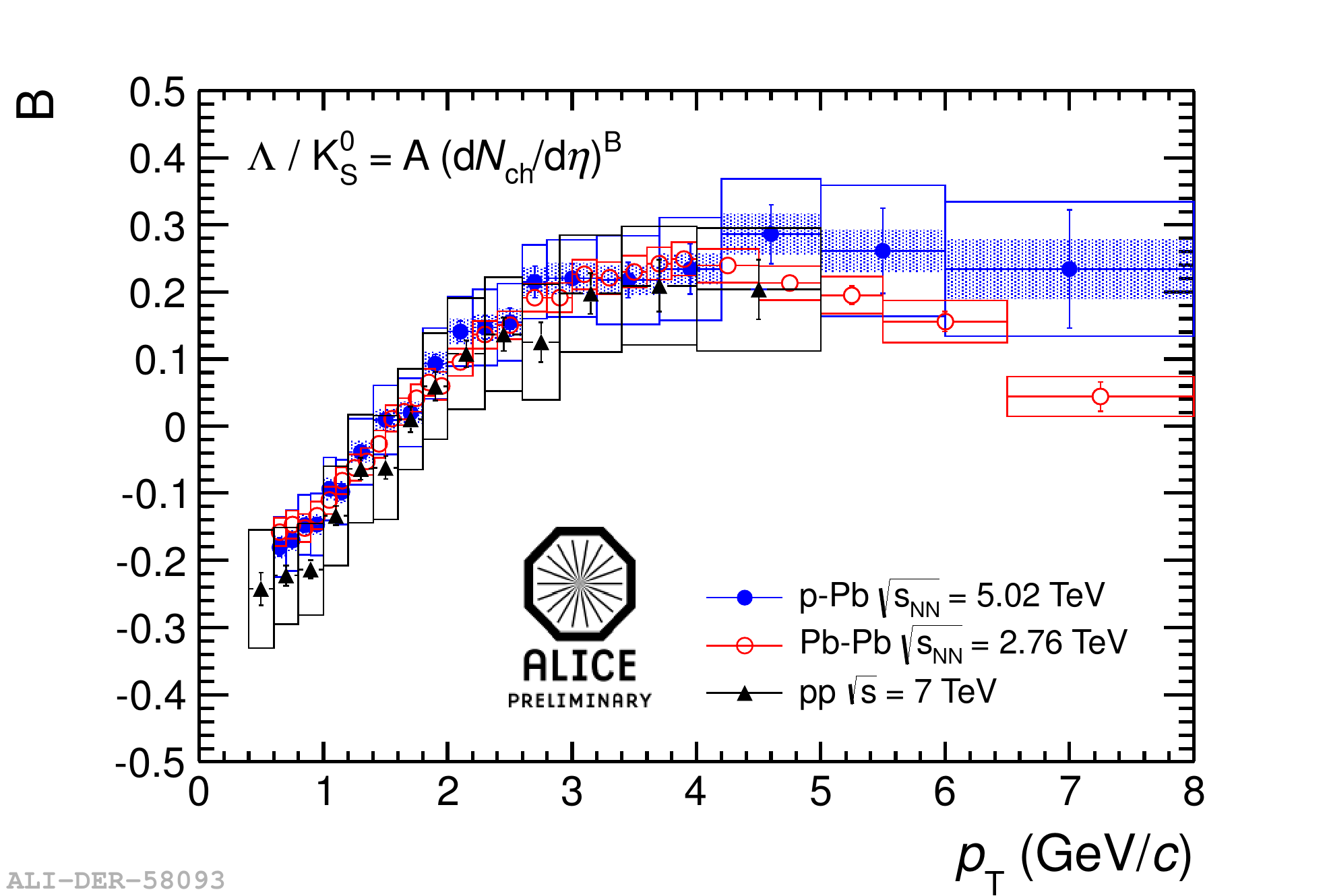}
  \caption{Exponent of the p/$\pi$ (left) and $\Lambda$/K$_{\rm S}^{0}$ (right) power-law fit as a function of $p_{\rm T}$ in p--Pb, Pb--Pb and pp collisions (pp only shown for $\Lambda$/K$_{\rm S}^{0}$).}
  \label{fig:power}
\end{figure}

The \pt\ distributions of \allpart\ in $0 < \ycms\ < 0.5$ are reported in~\cite{Abelev:2013haa} for different multiplicity intervals.
Particle/antiparticle as well as charged/neutral kaon transverse momentum distributions are identical within systematic uncertainties.
The \pt\ distributions show a clear evolution, becoming harder as the
multiplicity increases. The multiplicity dependence of the \pt\ spectral shape is stronger for heavier particles, as
 evident when looking at the ratios
\kpi~=~(\kap + \kam)/(\pip + \pim),  \ppi~=~(p + \pbar)/(\pip +
\pim) and  \lmb/\kzero\ as functions of
\pt, shown in Fig.~\ref{fig:ratios} for the 0--5\% and 60--80\%
event classes.  The ratios \ppi\ and \lmb/\kzero\ show
a significant enhancement at intermediate \pt~$\sim 3$~\gevc,
qualitatively reminiscent of the one measured in \PbPb\
collisions~\cite{prl-spectra, Abelev:2013vea, ALICE:2013xaa}.  The latter
is generally discussed in terms of collective flow or quark
recombination \cite{Fries:2003vb, Bozek:2011gq,Muller:2012zq}.
 A similar enhancement of the \ppi\ ratio in
high-multiplicity d--Au collisions has also been reported for RHIC
energies~\cite{Adare:2013esx}.

It is worth noticing that the ratio \ppi\ as a function of \dNdeta\ in
a given \pt-bin follows a power-law behavior: $\frac{\rm p}{\pi}\left(\pt\right) = A(\pt)
\times \left[\dNdeta\right]^{B(\pt)}$. As shown in Fig.~\ref{fig:scaling},
the same trend is also observed in \PbPb\ collisions. The exponent of
the power-law function exhibits the same value in both
collision systems (Fig.~\ref{fig:power}, left). The same feature
is also observed in the \lmb/\kzero\ ratio and this also holds in pp collisions (Fig.~\ref{fig:power},
right).

\section{Discussion}

\begin{figure}[t!]
  \centering
  \includegraphics[width=0.7\textwidth]{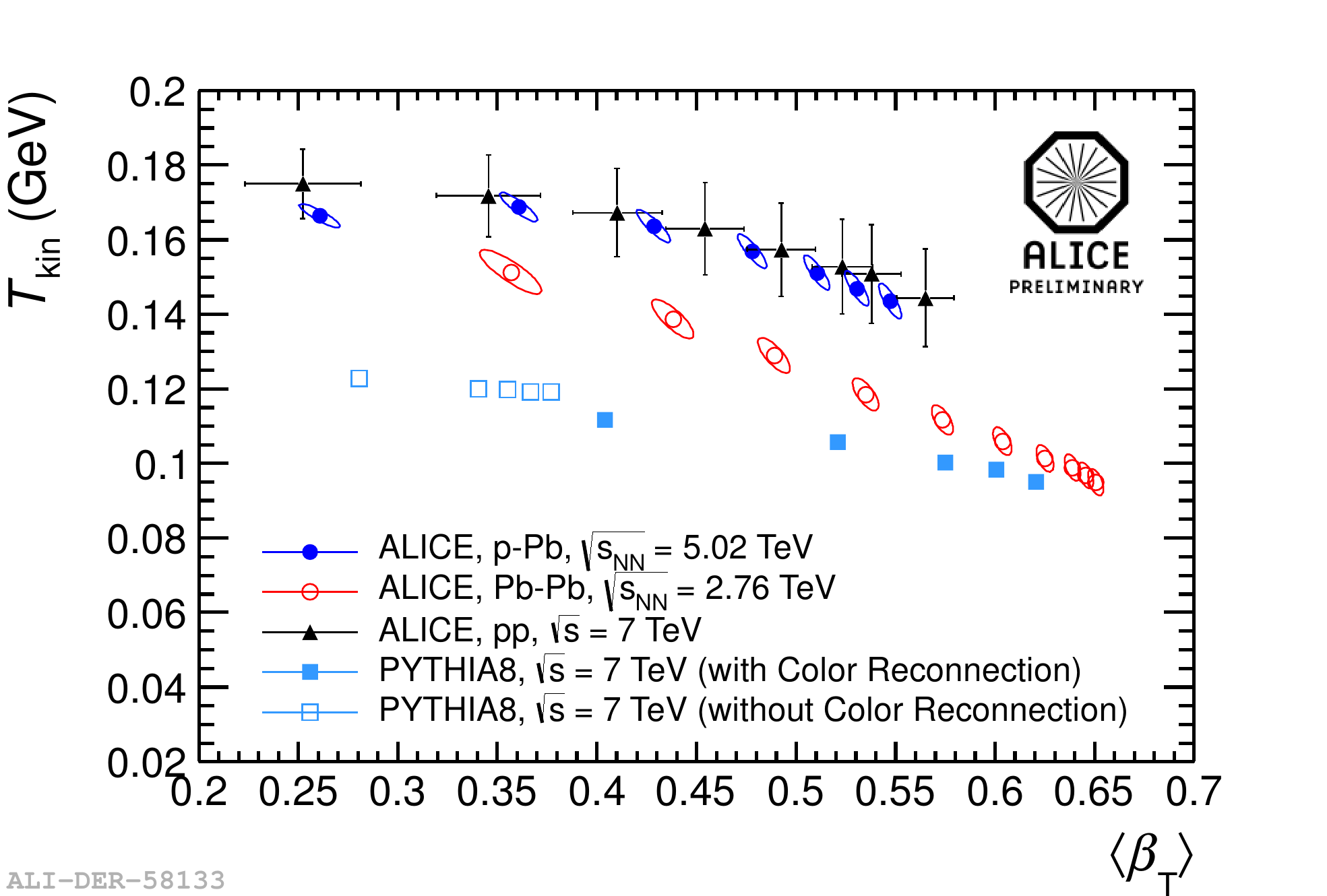}
  \caption{Results of blast-wave fits, compared to Pb--Pb data, pp data and MC simulations from PYTHIA8 with and without color reconnection. Charged-particle multiplicity increases from left to right. Uncertainties from the global fit are shown as correlation ellipses for p--Pb and Pb--Pb data and with errors bars for pp data.}
  \label{fig:blast-wave}
\end{figure}

\begin{figure}[t!]
  \centering
  \includegraphics[width=0.7\textwidth]{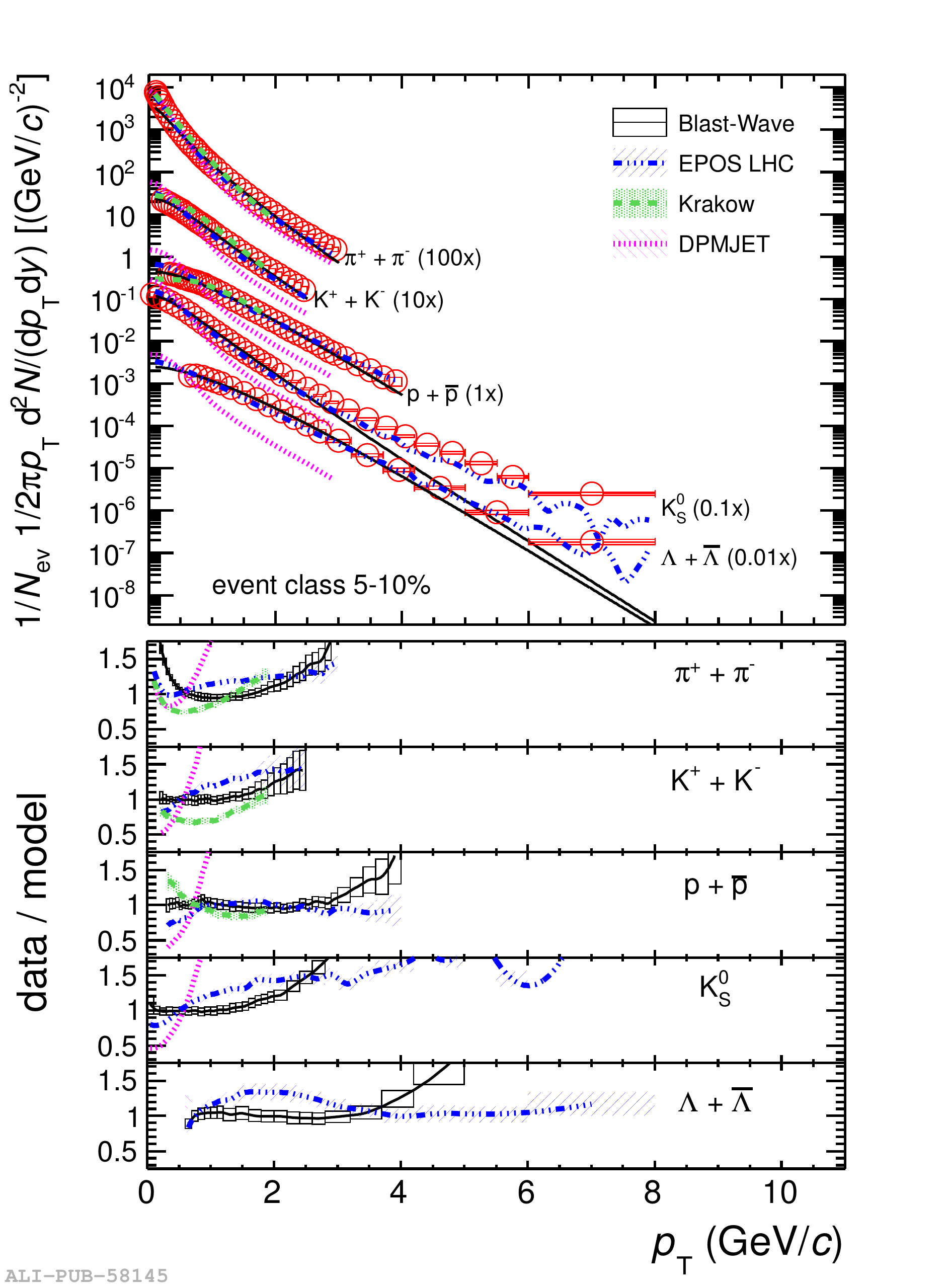}
  \caption{Pion, kaon, and proton transverse momentum distributions in the 5-10\% multiplicity class compared to the several models (see text for details).}
  \label{fig:BozekComparison}
\end{figure}

In heavy-ion collisions, the flattening of transverse momentum distribution and its
mass ordering find their natural explanation in the collective radial
expansion of the system~\cite{Heinz:2004qz}. This picture can be
tested in a blast-wave model~\cite{Schnedermann:1993ws} with a simultaneous fit to all
particles. This parameterization assumes a
locally thermalized medium, expanding collectively with a common
velocity field and undergoing an instantaneous common freeze-out.
The fit presented here is performed in the same range as 
in~\cite{prl-spectra, Abelev:2013vea}, also including \kzero\
and \lmb(\almb). The results are reported Fig.~\ref{fig:blast-wave}. 
Variations of the fit range lead to
large shifts ($\sim 10\%$) of the fit results (correlated across
centralities), as discussed for \PbPb\ data in~\cite{prl-spectra,  Abelev:2013vea}.
As can be seen in Fig.~\ref{fig:blast-wave}, the parameters show a
similar dependency with event multiplicity as observed with the \PbPb\ data. Within the limitations of
the blast-wave model, this observation is consistent with the presence
of radial flow in \pPb\ collisions. Under the assumptions of
a collective hydrodynamic expansion, a larger radial velocity in \pPb\
collisions has been suggested as a consequence of stronger radial
gradients in~\cite{Shuryak:2013ke}. On the other hand it is worth noticing that very similar results are obtained when performing the same study on pp spectra measured as a function of the event multiplicity. 
Other processes not related to hydrodynamic collectivity could also 
be responsible for the observed results.
This is illustrated in
Fig.~\ref{fig:blast-wave}, which shows the results obtained by
applying the same fitting procedure to transverse momentum distributions from the simulation of
pp collisions at \s~=~7 TeV with the PYTHIA8 event generator (tune
4C)~\cite{Corke:2010yf}, a model not including any collective system
expansion. The fit results are shown for PYTHIA8 simulations performed both with
and without the color reconnection
mechanism~\cite{Skands:2007zg,Schulz:2011qy}. With color
reconnection the evolution of PYTHIA8 transverse momentum distributions follows a
similar trend as the one observed for p--Pb, pp and Pb--Pb collisions at
the LHC, while without color reconnection it is not as strong.  This
generator study shows that other final state mechanisms, such as color
reconnection, can mimic the effects of radial flow~\cite{Ortiz:2013yxa}.

The \pt\ distributions in the 5-10\% bin are compared in
Fig.~\ref{fig:BozekComparison} with calculations from the DPMJET~\cite{Roesler:2000he},
Krak\'ow~\cite{Bozek:2011if} and EPOS LHC 1.99
v3400~\cite{Pierog:2013ria} models. 
The transverse momentum distributions in the 5-10\% multiplicity class
are compared to the predictions by Krak\'ow for $11 \leq N_{\rm part}
\leq 17$, since the \dNdeta\ from the model matches best with the
measured value in this class.  DPMJET and EPOS events have been
selected according to the charged particle multiplicity in the
\VZEROA\ acceptance in order to match the experimental selection.
DPMJET distributions are softer than the measured ones and the model
overpredicts the production of all particles for \pt\ lower than about
0.5--0.7 \gevc\ and underpredicts it at higher momenta.  At high-\pt,
the \pt\ spectra shapes of pions and kaons are rather well reproduced
for momenta above 1 and 1.5~\gevc\, respectively. Final state effects
may be needed in order to reproduce the data.  In fact, The Krak\'ow
model reproduces reasonably well the shape of pions and kaons below
transverse momenta of 1 \gevc\ where hydrodynamic effects are expected
to dominate. For higher momenta, the observed deviations for pions and
kaons could be explained in a hydrodynamic framework as due to the
onset of a non-thermal component. EPOS can reproduce the pion and
proton distributions within 20\% over the full measured range, while
larger deviations are seen for kaons and lambdas. It
is interesting to notice that when final state interactions are
disabled in EPOS, the description of many \pp\ and \pPb\ observables
worsens significantly~\cite{Pierog:2013ria}.

\section{Conclusions}

We presented a comprehensive measurement of \allpart\ in
\pPb\ collisions at \snn~=~5.02~TeV at the LHC. 
The transverse momentum distributions show a clear evolution
with multiplicity, similar to the pattern observed in high-energy pp and heavy-ion
collisions, where in the latter case the effect is usually attributed
to collective radial expansion.
Models incorporating final state effects give a better description of the data. 

\bibliographystyle{h-physrev}
\bibliography{biblio}

\end{document}